\documentclass[twocolumn,9pt]{article} 

\usepackage[square,numbers,sort&compress,comma]{natbib}

\usepackage{amsmath}
\usepackage{amssymb}
\usepackage{caption}
\usepackage{subcaption}
\usepackage{graphicx}
\usepackage{latexsym}
\usepackage{times}
\usepackage[pagewise]{lineno}
\usepackage{hyperref}

\usepackage[version=4]{mhchem}
\usepackage{chemformula}

\def\review#1{{\color{black}#1}}
\def\highlight#1{{\color{black}#1}}

\topmargin - 12pt 
\renewenvironment{abstract}%
              {
               \small
               {\bfseries \abstractname}
               \par
               \vspace{10pt}
              }

\renewcommand\abstractname{Abstract}

\newcommand{\nomenclature}
              [1]
              {
               \bgroup
               \flushleft
               \small\bf
               #1
               \par
               \egroup
              }

\renewcommand{\section}
              [1]
              {
               \bgroup
               \flushleft
               \small\bf
               \refstepcounter{section}
               \arabic{section}. #1
               \par
               \egroup
              }

\renewcommand{\subsection}
              [1]
              {
               \bgroup
               \flushleft
               \small\em
               \refstepcounter{subsection}
               \arabic{section}.
               \arabic{subsection}. #1
               \par
               \egroup
              }

\renewcommand{\subsubsection}
              [1]
              {
               \bgroup
               \flushleft
               \small\em
               \refstepcounter{subsubsection}
               \arabic{section}.
               \arabic{subsection}.
               \arabic{subsubsection}. #1
               \par
               \egroup
              }

  \newcommand{\acknowledgement}
              [1]
              {
               \bgroup
               \flushleft
               \small\bf
               #1
               \par
               \egroup
              }

  \newcommand{\sectionbib}
              [1]
              {
               \bgroup
               \flushleft
               \small\bf
               #1
               \par
               \egroup
              }

\begin{document}



\small
\baselineskip 10pt

\setcounter{page}{1}
\title{\LARGE \bf Direct numerical simulation of NOx formation in turbulent lean premixed hydrogen--air flames under engine-relevant conditions}

\author{{\large Chao Xu$^{*}$, Yiqing Wang, Riccardo Scarcelli}\\[10pt]
        {\footnotesize \em Transportation and Power Systems Division, Argonne National Laboratory, Lemont 60439, USA}\\[-5pt]
}

\date{}  

\twocolumn[\begin{@twocolumnfalse}
\maketitle
\rule{\textwidth}{0.5pt}
\vspace{-5pt}

\begin{abstract} 
In this study, direct numerical simulations (DNS) are employed to investigate NOx formation in turbulent lean premixed hydrogen--air flames under engine-relevant conditions. Various turbulence intensities and molecular transport models are examined to isolate the individual impacts of turbulence intensity, Lewis number, and preferential diffusion on local and global NO production. Results show that global NO production is significantly enhanced in all of the turbulent cases, reaching approximately five times the value of the 1D steady flame at a mixture residence time of 0.1~ms. Increasing turbulence intensity is found to have three competing effects on NO formation: (1) it strengthens turbulence--instability interactions by inducing local super-adiabatic hot spots and elevating the concentrations of key flame radicals within the flame brush, thereby promoting the NO reaction rate locally; (2) it accelerates the turbulent flame speed, reducing the flame-brush residence time and thus suppressing NO production globally; and (3) it reduces post-flame temperature fluctuations, suppressing thermal NO enhancement in the post-flame zone. \highlight{As a result, the global NO production is slightly lower at higher turbulence intensities among all the turbulent cases considered.} Lewis number effects are identified as the primary mechanism driving thermodiffusive NO enhancement, with preferential diffusion playing a secondary role, as evidenced by the nearly identical \review{mean profiles of the NO reaction rate between unity Lewis number turbulent flames and their 1D steady flame counterparts in both the progress variable space and the residence time space}. Finally, an excellent correlation between the peak conditional mean NO reaction rate and the stretch factor is identified, and a conceptual model is proposed to \review{improve the predictions of global NO production} in practical engine simulations. The findings highlight that turbulence--chemistry interaction is critical for accurately predicting NOx formation in thermodiffusively unstable hydrogen flames.
\end{abstract}

\vspace{5pt}

{\bf Novelty and significance statement}

\vspace{10pt}
This work presents the first systematic investigation of the individual impacts of turbulence intensity, Lewis number effects, and preferential diffusion on NOx formation in turbulent lean premixed hydrogen flames under engine-relevant conditions. We report that NO can increase by a factor of five in turbulent flames relative to the 1D steady flame, and demonstrate, for the first time, that as turbulence intensity increases, the local NO production rate within the turbulent flame brush increases monotonically, while the overall NO level decreases due to the shortened flame-brush residence time \highlight{and the suppressed temperature fluctuations in the post-flame zone}. We further show that Lewis number effects are the primary driver of NO enhancement, while preferential diffusion plays a secondary role. These findings address a longstanding challenge in engine NOx prediction and provide physical insights to guide model development for ultra-lean, high-pressure hydrogen combustion applications. 

\vspace{5pt}
\parbox{1.0\textwidth}{\footnotesize {\em Keywords:} Turbulent premixed flame; Hydrogen combustion; Direct numerical simulation; NOx; Engine-relevant conditions}
\rule{\textwidth}{0.5pt}
*Corresponding author.
\vspace{5pt}
\end{@twocolumnfalse}] 

\section{Introduction\label{sec:introduction}} \addvspace{5pt}

Hydrogen (\ch{H2}) is a promising alternative fuel for heavy-duty transport, power generation, and various industrial processes. Accurate prediction of nitrogen oxides (NOx) formation from combustion systems such as internal combustion engines and gas turbines is crucial to engine design and optimization. Current models rely on the widely used Zel'dovich \review{model}~\cite{heywood1988combustion} and its variants, or on detailed NOx chemistry when reaction pathways other than thermal NOx become important~\cite{Glarborg2018a}. These approaches have shown good predictive capability for conventional engines operating on natural gas, gasoline, and diesel fuels~\cite{2012Mobasheri,2020Schluckner,2023Fabio}.

More recently, these models have been extended to hydrogen engines, showing adequate capability for predicting NOx~\cite{knop2008modelling,rakopoulos2011combined,zhu2019research,lu2024numerical,sfriso2024combination}. Most existing studies have focused on light-duty operation under stoichiometric or moderately lean conditions (e.g., with an equivalence ratio of $\phi > 0.5$). Due to the high propagation speed and the high adiabatic flame temperature of hydrogen flames, practical heavy-duty engines typically operate under ultra-lean conditions to avoid knock while mitigating NOx, with $\phi$ as low as 0.3. However, NOx modeling efforts for such operating conditions remain scarce, \review{particularly in the Reynolds-averaged Navier-Stokes (RANS) framework}. One notable exception is the study by Maio et al.~\cite{Maio2022} on a heavy-duty \ch{H2} spark-ignition engine at $\phi = 0.4$--0.55. Although their simulations correctly captured the trend of NOx with changing $\phi$, the magnitudes of NO and \ch{NO2} were under-predicted by a factor of 2--3 and 2--5, respectively. A more recent study by Scarcelli et al.~\cite{scarcelli2025} on a large-bore off-road engine operating under ultra-lean conditions ($\phi = 0.3$--0.4) reported a similar or larger underprediction of total NOx (by a factor of 3--12) despite good agreement with experimental data for the main combustion process. They also showed that the prediction error in NOx increases as the mixture becomes leaner. These results clearly highlight a research gap in better understanding NOx formation mechanisms for ultra-lean turbulent hydrogen combustion under engine-relevant conditions.

Direct numerical simulations (DNS) offer a unique approach for investigating fundamental flame properties and NOx formation in turbulent flames. Nevertheless, literature on DNS studies of NOx formation in turbulent premixed hydrogen flames is notably sparse, with only a few studies available. Day et al.~\cite{Day2011} simulated NOx formation in lean premixed turbulent \ch{H2}/\ch{O2}/\ch{N2} flames using two-dimensional DNS, finding that localized hot spots led to significant NOx production. They also showed that this NOx production enhancement increases and the dominant pathway shifts from \ch{N2O} to NNH as $\phi$ increases from 0.3 to 0.4. Bell et al.~\cite{Bell2013} further simulated NOx formation in a low-swirl stabilized lean premixed \ch{H2}--air flame, demonstrating that cellular burning structures provide unique NOx production sites that lead to both local and global NOx enhancement. Fathi et al.~\cite{Fathi2025} investigated turbulent counterflow flames of lean \ch{H2}--air mixtures at different strain rates, observing that NOx formation was suppressed with increasing strain. Beyond turbulent conditions, several studies have examined NOx in laminar hydrogen flames, focusing on the effects of curvature and strain~\cite{Porcarelli2024,Wen2024b,Wen2024a}. They found that NOx was suppressed as strain rate increases~\cite{Porcarelli2024}, while NOx shows a strong positive correlation with flame curvature under thermodiffusively unstable conditions, especially at high pressure (8~bar)~\cite{Wen2024b}. Nevertheless, all these prior studies were conducted at relatively low pressures ($<8$~bar) and ambient temperature conditions, which differ substantially from practical engine environments.

In this study, we employ spectral element DNS to investigate NOx formation in turbulent premixed \ch{H2}--air flames under high-pressure, high-temperature, ultra-lean conditions. The primary focus is on the impact of turbulence and the individual roles of Lewis number effects (between heat and the deficient reactant) and preferential diffusion (between different species), as well as on the implications for NOx modeling in practical engine applications.

\section{Numerical methodologies\label{sec:sections}} \addvspace{5pt}

\begin{figure*}[h!]
\centering
\includegraphics[trim={0.2cm 5.5cm 0.2cm 5.5cm}, clip=true, width=1\linewidth]{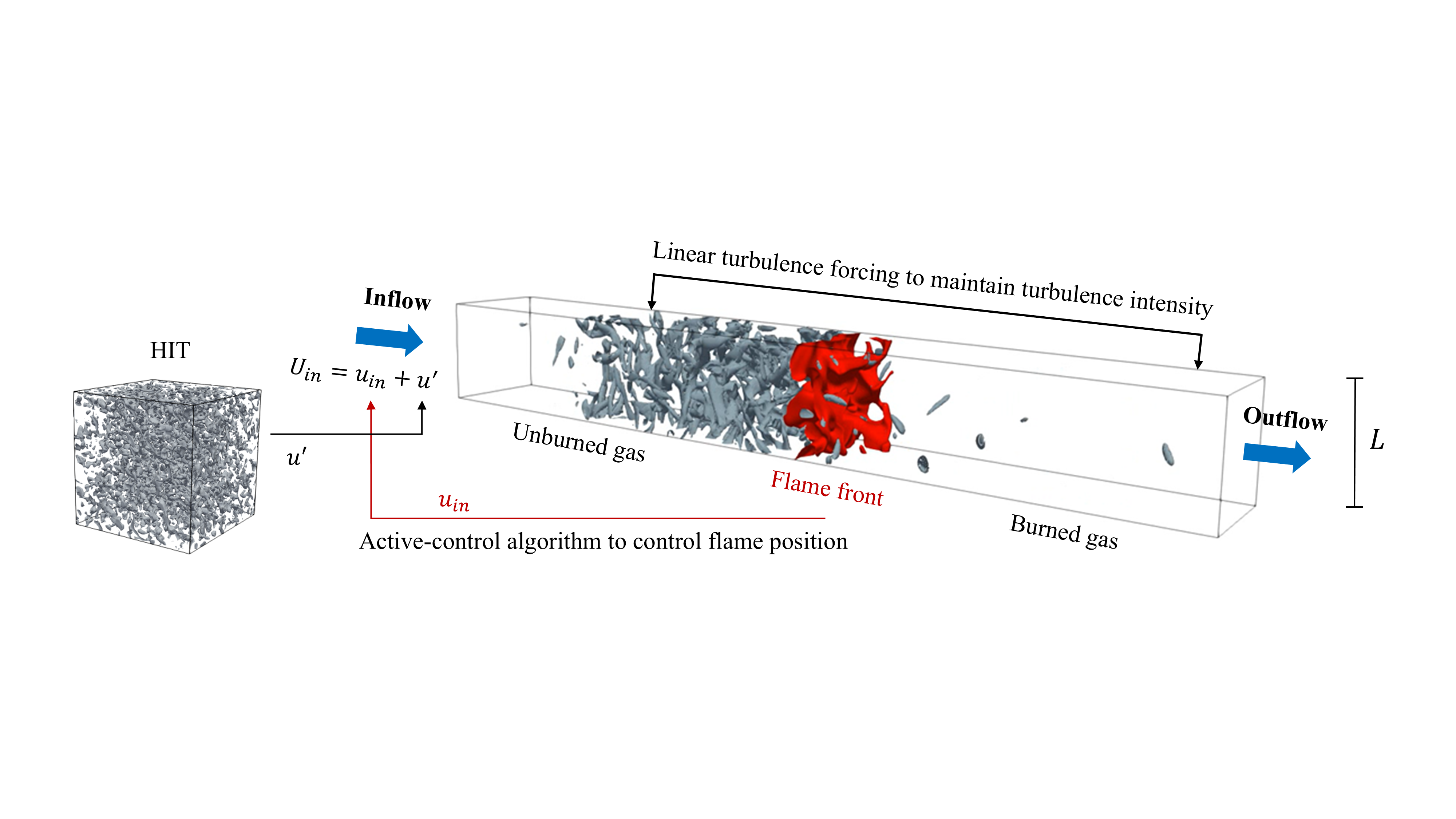}
\caption{\footnotesize DNS configuration. Gray represents vortex structure based on Q-criterion. Red represents the flame front.}
\vspace{-0.2 in}
\label{fig_DNS_configuration}
\end{figure*}

The DNS configuration is shown in Fig.~\ref{fig_DNS_configuration}, similar to that in~\cite{Aspden2017}. Statistically planar turbulent premixed flames are simulated using a doubly periodic inflow--outflow configuration for lean \ch{H2}/air mixtures at an equivalence ratio of $\phi = 0.35$, an unburned temperature of $T_u = 900$~K, and a pressure of $P = 100$~atm, representative of typical thermodynamic conditions in heavy-duty engines. Chemical kinetics are described using a 17-species hydrogen--NOx \review{reaction model} based on C3MechLite~\cite{Murakami2025} with helium and argon removed, \review{which is one of the newest detailed \ch{H2} reaction models with a built-in NOx sub-model and extensive experimental validation.} The adiabatic flame temperature is $T_b = 1831$~K. 

DNS is performed using the spectral element code Nek5000~\cite{NEK5000} in a low-Mach-number formulation. The code employs high-order spatial discretization, exhibits excellent scalability on modern computing platforms, and has been successfully applied to turbulent reacting flows~\cite{Xu2022a,Xu2023,Wang2025}. Time integration uses a semi-implicit second-order backward differentiation method with an operator-integrator factor scheme (OIFS), allowing variable time-stepping with CFL numbers up to 2. An iterative dynamic chemical stiffness removal method (IDCSR)~\cite{Xu2025a} is employed to accelerate chemistry integration.

Seven turbulent cases are considered, with varying turbulence parameters and molecular transport models as listed in Table~\ref{table_DNS_parameter}. \review{The number ``XX'' in the case name ``UXX'' denotes the value of $u'/S_L$.} The first four cases investigate the effects of turbulence intensity ($u'/S_L$) at a fixed turbulence length scale ratio ($l_T/l_f = 2$) in the thin reaction zones regime using the mixture-averaged molecular transport model. Following a methodology similar to Lee et al.~\cite{2021Lee}, the next three cases examine the effects of molecular transport modeling by considering: (1) unity Lewis number for all species (``U10-Le1''); (2) mixture-averaged diffusivity for \ch{H2} only, with all other species assigned the same diffusivity as \ch{H2} (``U10-LeH$_2$''); and (3) mixture-averaged diffusivity ($D_{km}$) multiplied by $\alpha / D_{H_2}$ for all species (``U10-Le1Mult''), where $\alpha$ is thermal diffusivity. Effectively, ``U10-Le1'' eliminates both Lewis number and preferential diffusion effects, while ``U10-LeH$_2$'' and ``U10-Le1Mult'' retain Lewis number and preferential diffusion effects, respectively. \review{An eighth case, a freely propagating 3D laminar flame denoted as ``3D-Laminar'', is also simulated using the same computational domain as U05-U20.}
Soret and Dufour effects are neglected in all cases.

\begin{table}[ht] 
\vspace{-0.05 in}
\footnotesize
\centering
\caption{Simulation parameters for the DNS cases. $S_L$ is laminar flame speed in m/s, $l_f$ is laminar flame thickness in $\mu$m, $u'$ is turbulence fluctuation velocity, $l_T$ is the integral length scale, Karlovitz number is defined as Ka~$=(l_T/l_f)^{-0.5}(u'/S_L)^{1.5}$. }
\begin{tabular}{cccccc}
\hline
Case & $S_L$ & $l_f$ & $u'/S_L$ & $l_T/l_f$ & Ka \\
\hline
U05     & 0.23 & 19 & 5 & 2 & 8 \\
U10     & 0.23 & 19 & 10 & 2 & 22 \\
U15     & 0.23 & 19 & 15 & 2 & 41 \\
U20     & 0.23 & 19 & 20 & 2 & 63 \\
U10-Le1      & 0.51 & 9.4 & 10 & 2 & 22 \\
U10-LeH$_2$  & 0.23 & 21 & 10 & 2 & 22 \\
U10-Le1Mult  & 0.58 & 8 & 10 & 2 & 22 \\
\review{3D-Laminar}  & 0.23 & 19 & - & - & - \\
\hline 
\end{tabular}
\vspace{-0.1 in}
\label{table_DNS_parameter}
\end{table}

The computational domain has dimensions $L \times 8L \times L$ (except for U20 which has dimensions $L \times 12L \times L$). Inflow and outflow boundary conditions are imposed in the streamwise direction ($y$), while periodic boundary conditions are applied in the lateral directions. The solution from a one-dimensional (1D), steadily-propagating, unstretched laminar premixed flame \highlight{(referred to as ``1D steady flame'' hereafter)} is mapped along the $y$-direction with the flame front positioned at the domain center. The flow field is initialized with homogeneous isotropic turbulence (HIT) superimposed on the 1D steady flame profile. An active-control algorithm~\cite{2007Bell} maintains the mean flame position at $y = 4L$. Linear turbulence forcing~\cite{2016Bobbitt} is applied in the unburned gas region ($y = 0.5L$ to $3.6L$) to sustain the desired turbulence level. This forcing scheme establishes a turbulent integral length scale proportional to the domain width $L$, such that $l_T = 0.19L$~\cite{2016Bobbitt}. In all DNS cases, spectral elements are uniformly distributed with seventh-order polynomials (eight grid points per element in each spatial direction) and 24 spectral elements spanning $L$. This provides 16 grid points per laminar flame thickness and ensures $k_{max}\eta > 2$, where $\eta$ is the Kolmogorov length scale and $k_{max}$ is the maximum resolved wavenumber magnitude.

\section{Results and discussion\label{sec:figtabeqn}} \addvspace{5pt}

\review{All DNS statistics are collected over a sufficiently long duration ($\sim30 \tau_e$ where $ \tau_e=l_T/u'$ is the eddy turn over time) after the flame has reached a statistically steady state.} Turbulent flame speed $S_T$ normalized by $S_L$, flame surface area $A_T$ normalized by $A_L$, stretch factor $I_0=(S_T/S_L)/(A_T/A_L)$, and flame brush thickness $\delta_b$ normalized by $l_f$ from all DNS cases are reported in Table~\ref{table_ST_AT_I0}. Here, \review{$A_T$ is computed based on the generalized flame surface density ($\Sigma=|\nabla c|$) according to Ref.~\cite{Wang2025}, where $c$ is the progress variable defined from the \ch{H2} mass fraction.} $A_L=L\times L$ is the projected flame area in the streamwise direction. The flame brush thickness $\delta_b$ is computed \review{using an integral method (see supplementary material)} based on the temporally and planarly averaged one-dimensional temperature profile. As expected, increasing turbulence intensity leads to enhanced turbulent flame speed, flame surface area, and stretch factor, consistent with the authors' previous findings~\cite{Wang2025}. The turbulent flame speeds for Cases ``U10-Le1'' and ``U10-Le1Mult'' are significantly lower, with $I_0$ close to unity, while Case ``U10-LeH$_2$'' remains comparable to Case ``U10'', consistent with the literature~\cite{Lee2022}. \highlight{Note that the sensitivity of the results to the choice of $A_T$ definition merits further investigation.} The following sections examine the effects of turbulence, Lewis number, and preferential diffusion on NOx formation. \review{We mainly focus on NO, as the \ch{NO2} mass fraction is one order of magnitude lower than that of NO for the conditions considered in this study.} 

\begin{table}[ht] 
\scriptsize 
\centering
\caption{Turbulent flame speed, flame surface area, stretch factor, and flame brush thickness from DNS, all normalized by the 1D steady flame counterparts.}
\begin{tabular}{ccccc}
\hline
Case & $S_T/S_L$ & $A_T/A_L$ & $I_0$ & $\delta_b/l_f$ \\
\hline
U05          & 13.1 & 2.94 & 4.44 & 5.13 \\
U10          & 25.2 & 4.28 & 5.88 & 5.72 \\
U15          & 34.3 & 4.93 & 6.97 & 6.12 \\
U20          & 49.2 & 6.31 & 7.80 & 7.26 \\
U10-Le1      & 3.55 & 3.64 & 0.97 & - \\
U10-LeH$_2$  & 24.4 & 3.74 & 6.52 & - \\
U10-Le1Mult  & 2.55 & 3.31 & 0.76 & - \\
\review{3D-Laminar}   & 2.43 & 1.20 & 2.03 & 2.19 \\
\hline 
\end{tabular}
\vspace{-0.1 in}
\label{table_ST_AT_I0}
\end{table}

\subsection{Effects of turbulence\label{sec:turbulenceEffect}} \addvspace{10pt}

\begin{figure*}[ht!]
\centering
\includegraphics[trim={0.2cm 4cm 0.2cm 8cm}, clip=true, width=0.8\linewidth]{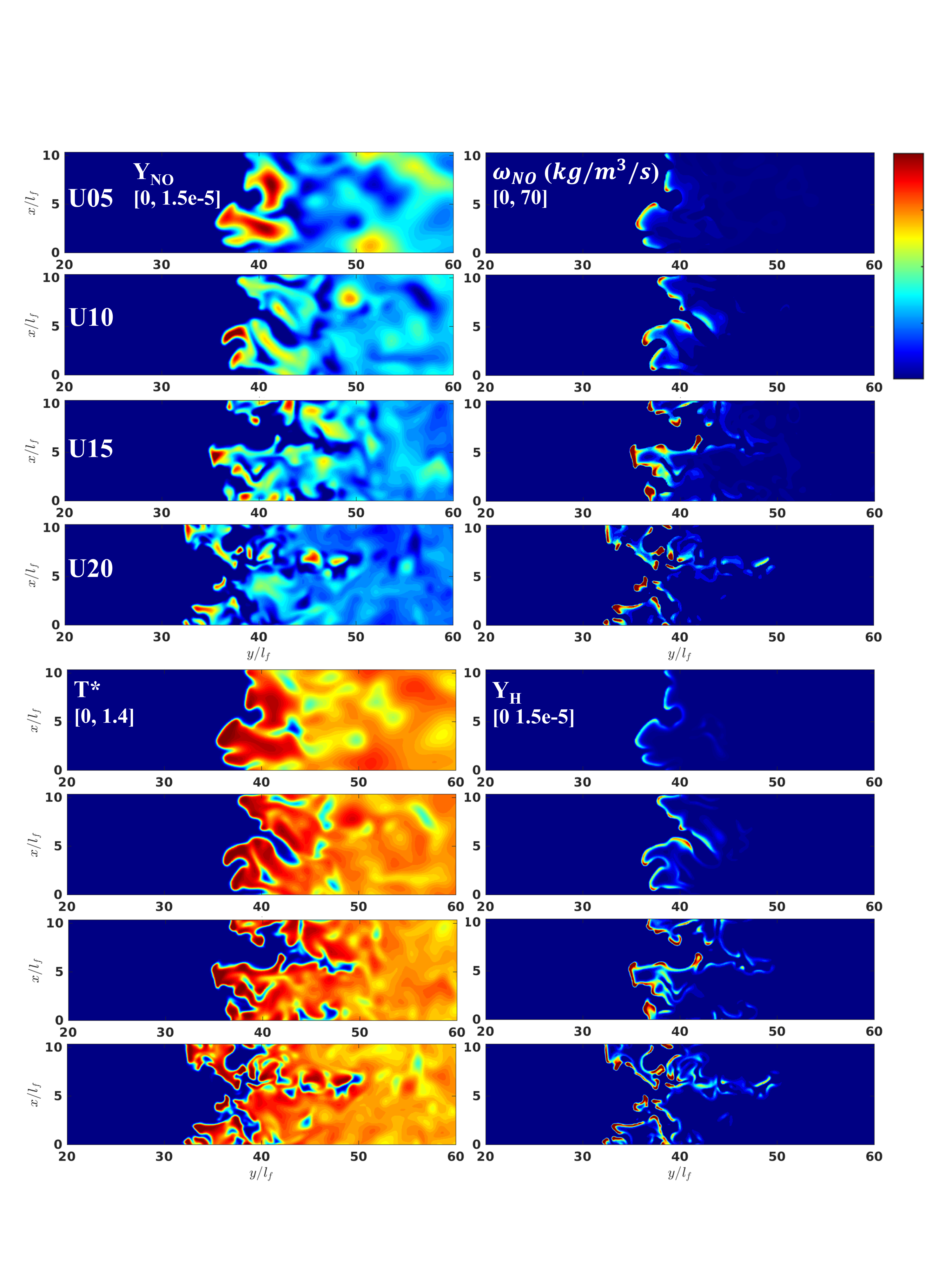}
\caption{\footnotesize Contour plots of $Y_{NO}$, $\omega_{NO}$, normalized temperature $T^*$, and $Y_{H}$ at different turbulence intensities.}
\vspace{-0.1 in}
\label{fig_2Dcountours_major_diffU}
\end{figure*}

\begin{figure*}[ht!]
\centering
\includegraphics[trim={1cm 1cm 1cm 1cm}, clip=true, width=1.0\linewidth]{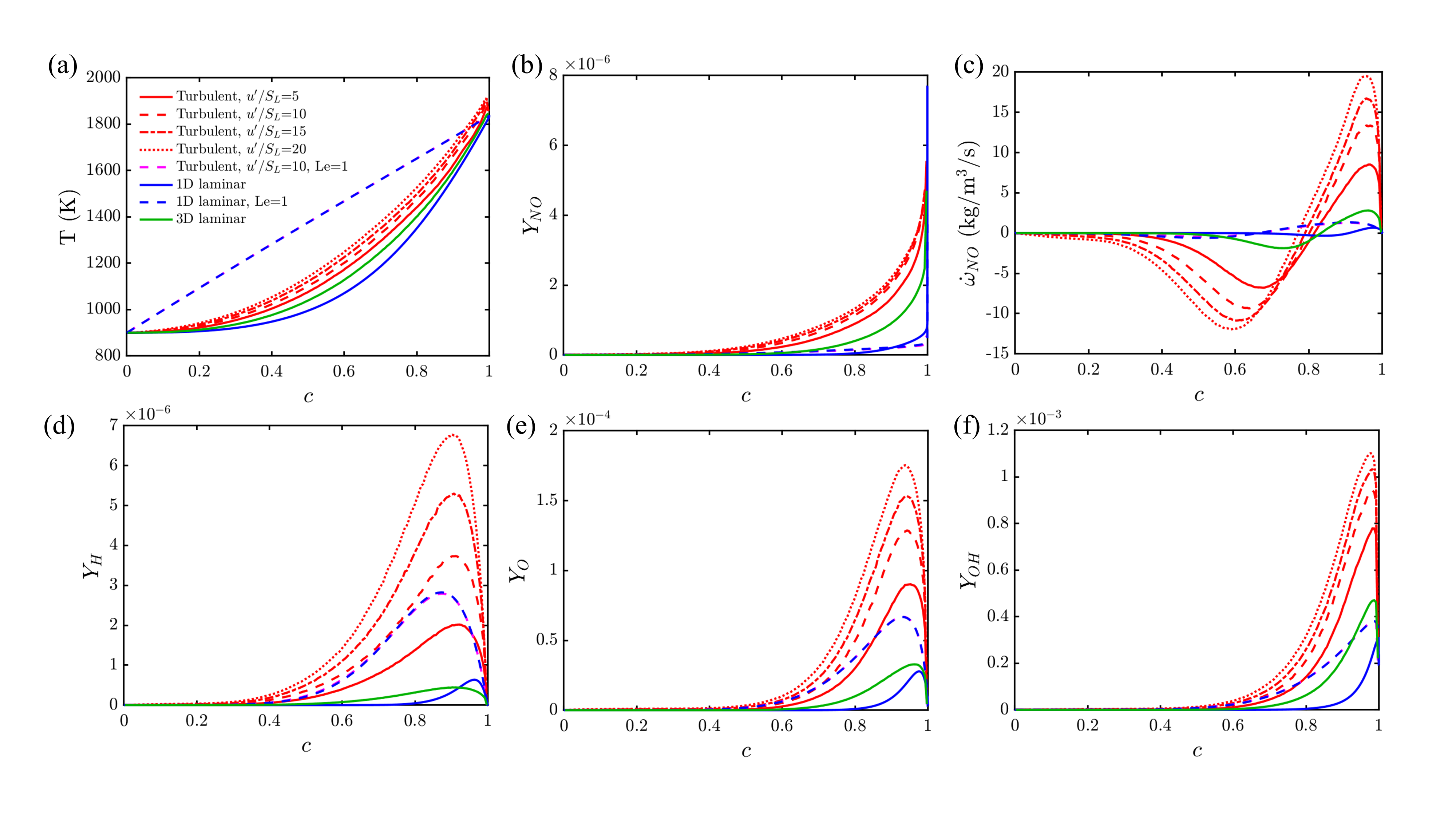}
\caption{\footnotesize Mean profiles of $T$, $Y_{NO}$, $\dot{\omega}_{NO}$, $Y_{H}$, $Y_{O}$, and $Y_{OH}$ conditioned on progress variable, at different turbulence intensities.}
\vspace{-0.1 in}
\label{fig_scatter_plot}
\end{figure*}

Figure~\ref{fig_2Dcountours_major_diffU} illustrates the spatial distributions of NO mass fraction ($Y_{NO}$) and its source term ($\dot{\omega}_{NO}$), as well as the basic flame structure in terms of normalized temperature $T^* = (T - T_u)/(T_b - T_u)$ and H mass fraction ($Y_{H}$), at different turbulence intensities. As expected, increasing turbulence intensity leads to a more wrinkled flame structure and smaller-sized cellular structures arising from thermodiffusive instability, manifested by super-adiabatic local hot spots with peak temperatures of $\sim$2200~K. This behavior is consistent with previous observations by Bell et al.~\cite{Bell2013}. Enhanced NO source terms coincide with elevated H radical levels at higher turbulence intensities. \review{Pockets of high NO mass fraction are} observed within the flame zone, generally co-located with local hot spots and regions of positive curvature. Interestingly, as turbulence intensity increases, NO appears to be suppressed at comparable downstream locations, a phenomenon that will be further discussed below. A strong spatial correlation exists between the NO source term and H radical distributions; although not shown, similar correlations are observed for other key radicals including O, OH, and \ch{HO2}.

Figure~\ref{fig_scatter_plot} presents the mean profiles of temperature, $Y_{NO}$, $\omega_{NO}$, and the mass fractions of H, O, and OH radicals conditioned on the progress variable ($c$). Here, $c = 1 - Y_{H_2}/Y_{H_2}^0$ with $Y_{H_2}^0$ being the unburned \ch{H2} mass fraction. The following discussion focuses on the non-unity Lewis number cases. The mean temperature profiles are broadly similar across the turbulent cases while remaining slightly elevated relative to both 1D and 3D laminar profiles. In contrast, \review{the NO mass fractions in all the turbulent cases are} one order of magnitude larger than those in the 1D steady flame for $c<1$, with \review{the values from the 3D laminar case lying in between. Higher turbulence intensity also yields a greater local NO mass fraction at progress variables well below unity}. This trend is confirmed by the $\omega_{NO}$ profiles, which exhibit a clear dependence of the peak NO reaction rate on turbulence intensity. Notably, increasing turbulence intensity enhances both the production and consumption rates of NO, a consequence of the elevated flame radical concentrations shown in Fig.~\ref{fig_scatter_plot}d--f. These results indicate that NO formation pathways within the flame brush can be strongly affected by turbulence-chemistry interaction. \review{In particular, the flame radicals (O, H, and OH) directly participate in the initiation and NO conversion reactions of these pathways, such as $\mathrm{O + N_2 = NO + N}$ and $\mathrm{N + OH = NO + H}$ (thermal route), $\mathrm{O + N_2(+M) = N_2O(+M)}$, $\mathrm{N_2O + O = 2NO}$ and $\mathrm{N_2O + H = NO + NH}$ (\ch{N2O} route), and $\mathrm{H + N_2(+M) = NNH(+M)}$ and $\mathrm{NNH + O = NO + NH}$ (NNH route)}. Note that while similar NO enhancement has been observed in 2D and 3D laminar flames of lean hydrogen mixture in the literature~\cite{Wen2024b,Wen2024a}, turbulence can further promote such effect as demonstrated in \review{Fig.~\ref{fig_scatter_plot}.}

\begin{figure}[ht!]
\centering
\includegraphics[trim={5.5cm 0 5.5cm 0.4cm}, clip=true, width=0.9\linewidth]{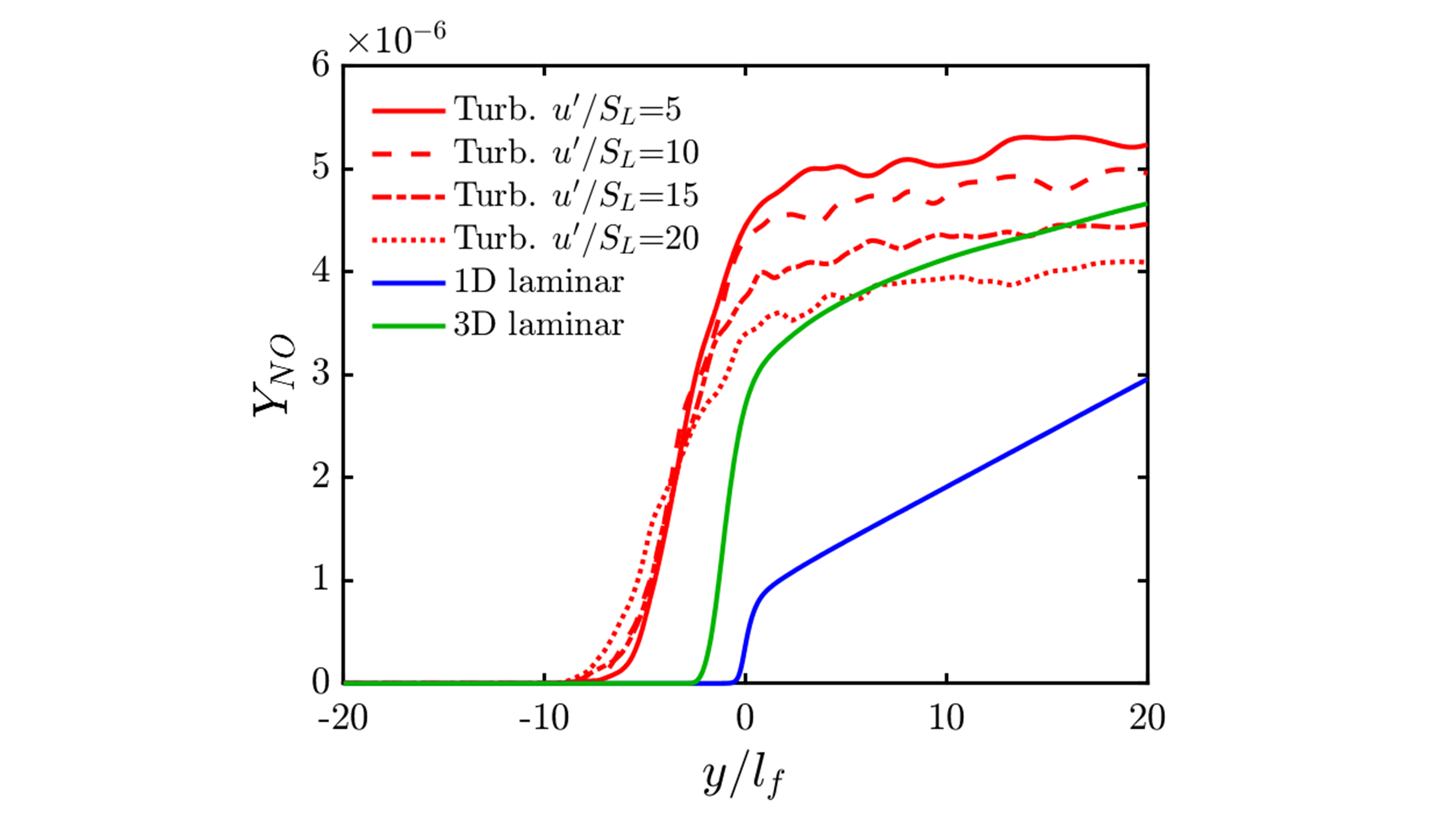}
\caption{\footnotesize Planar- and time-averaged spatial $Y_{NO}$ profiles at different turbulence intensities. The profiles are shifted by $4L$ so that all flames are centered at $y=0$.}
\vspace{-0.2 in}
\label{fig_YNO_spatialProfile_diffU}
\end{figure}

\begin{figure}[hb!]
\centering
\vspace{-0.2 in}
\includegraphics[trim={6cm -0.2cm 6cm 0.4cm}, clip=true, width=0.9\linewidth]{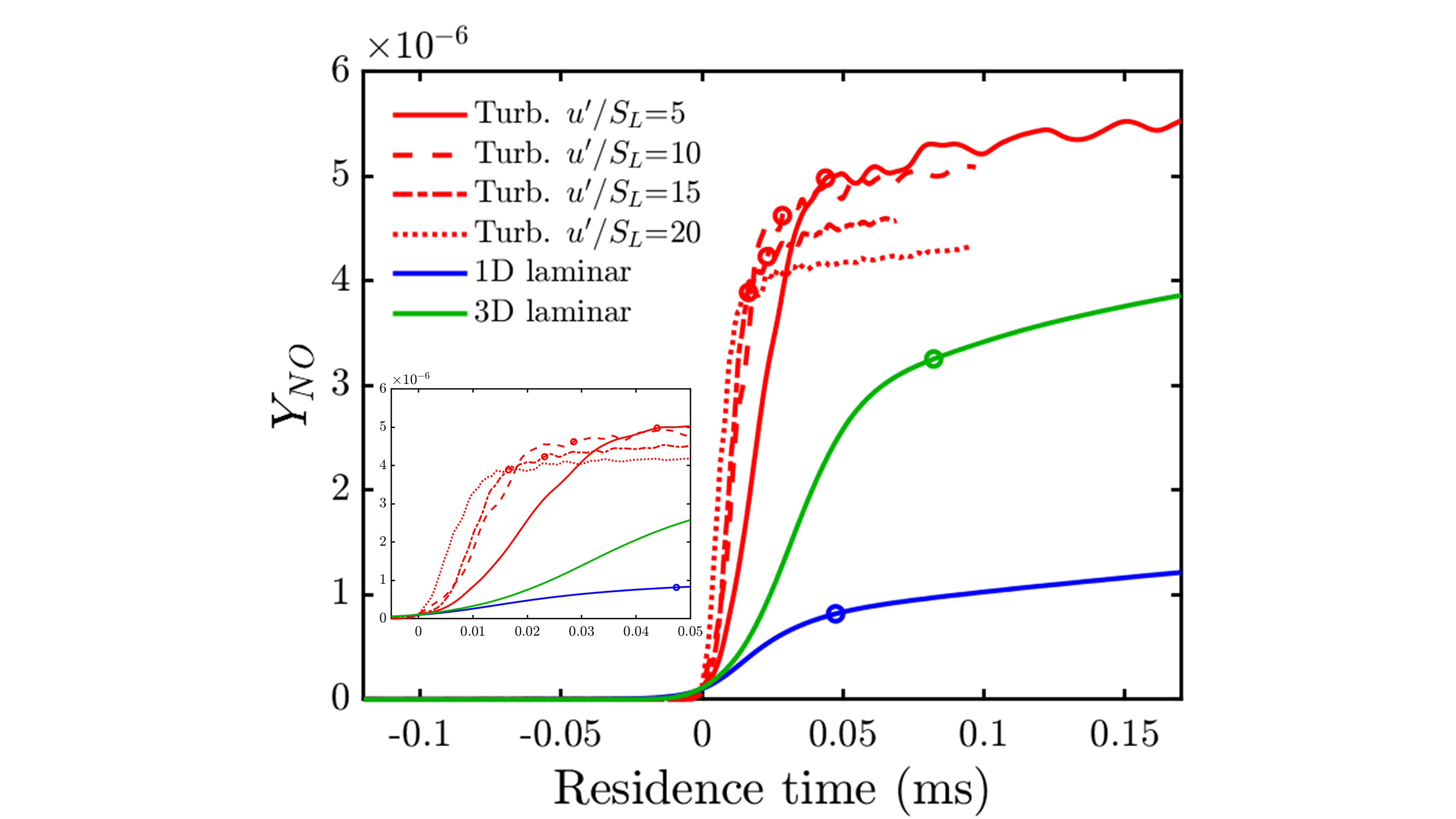}
\caption{\footnotesize Planar- and time-averaged $Y_{NO}$ profiles in the residence time space, at different turbulence intensities. Time zero is defined at $Y_{NO}=10^{-7}$, representing NO formation onset. Open circles \review{mark the end of the flame zone: $\tau_{res}^{flame}=0.044$~ms for U05, 0.029~ms for U10, 0.023~ms for U15, 0.017~ms for U20, 0.047~ms for the 1D steady flame, and 0.082~ms for the 3D laminar case.}}
\label{fig_YNO_residence_time_Profile_diffU}
\vspace{-0.1 in}
\end{figure}

Figure~\ref{fig_YNO_spatialProfile_diffU} shows spatial profiles of the mean NO mass fraction. At first glance, turbulence clearly enhances global NO formation relative to the 1D solution; however, increasing turbulence intensity \review{appears to reduce the spatial gradient of the NO profile and suppress the global NO formation. The reduction in the spatial gradient of NO} seemingly contradicts the turbulence-induced enhancement of the NO reaction rate observed in Fig.~\ref{fig_scatter_plot}. It should be noted, however, that spatial representations can be misleading, since flames with different turbulence intensities exhibit substantially different flame speeds and therefore different residence times.

It is therefore more instructive to examine the NO mass fraction as a function of \highlight{mean} mixture residence time ($\tau_{res}$), as shown in Fig.~\ref{fig_YNO_residence_time_Profile_diffU}. Here, $\tau_{res}$ is defined as $\tau_{res}=\int_{y_{NO-onset}}^{y} \frac{1}{u_y} dy$ where $u_y$ is the planar- and time-averaged streamwise velocity, $y_{NO-onset}$ the \textit{y}-location corresponding to NO formation onset ($Y_{NO} = 10^{-7}$). The flame zone starts at $Y_{NO} = 10^{-7}$ and ends at $c = 0.99$, i.e., 1\% unburnt \ch{H2}. \review{Even without turbulence, the thermodiffusive instability alone leads to significant NO enhancement already, as seen in the 3D laminar case. Adding turbulence further promotes this effect. In particular,} as turbulence intensity increases, the local NO production rate within the turbulent flame brush increases monotonically, a trend that cannot be inferred from the spatial profiles. The enhanced NO production rate within the flame brush at higher turbulence levels can be attributed to the increased occurrence of super-adiabatic hot spots observed in Fig.~\ref{fig_2Dcountours_major_diffU}, and is consistent with the elevated levels of key flame radicals shown in Fig.~\ref{fig_scatter_plot}. By inducing local hot spots of $\sim$2200~K, thermal NO formation rate is significantly accelerated, while the intensified radical pool can further promote the \ch{N2O} and NNH pathways. Once the mixture leaves the main flame brush and enters the post-flame zone ($c > 0.99$), NO continues to form but at a rate much closer to the 1D steady flame. Note that the equilibrium value $Y_{NO}^{eq} = 3.8 \times 10^{-3}$ for the condition considered here suggests that $Y_{NO}$ would continue to grow as residence time increases beyond the current computational domain. \highlight{While higher-order statistics of residence time could be obtained by tracking individual Lagrangian fluid particles, the mean residence time employed here is of primary interest, given the focus on RANS applications in this study.}

\begin{figure}[b!]
\centering
\vspace{-0.2 in}
\includegraphics[trim={4.5cm -0.5cm 6cm 0.8cm}, clip=true, width=0.9\linewidth]{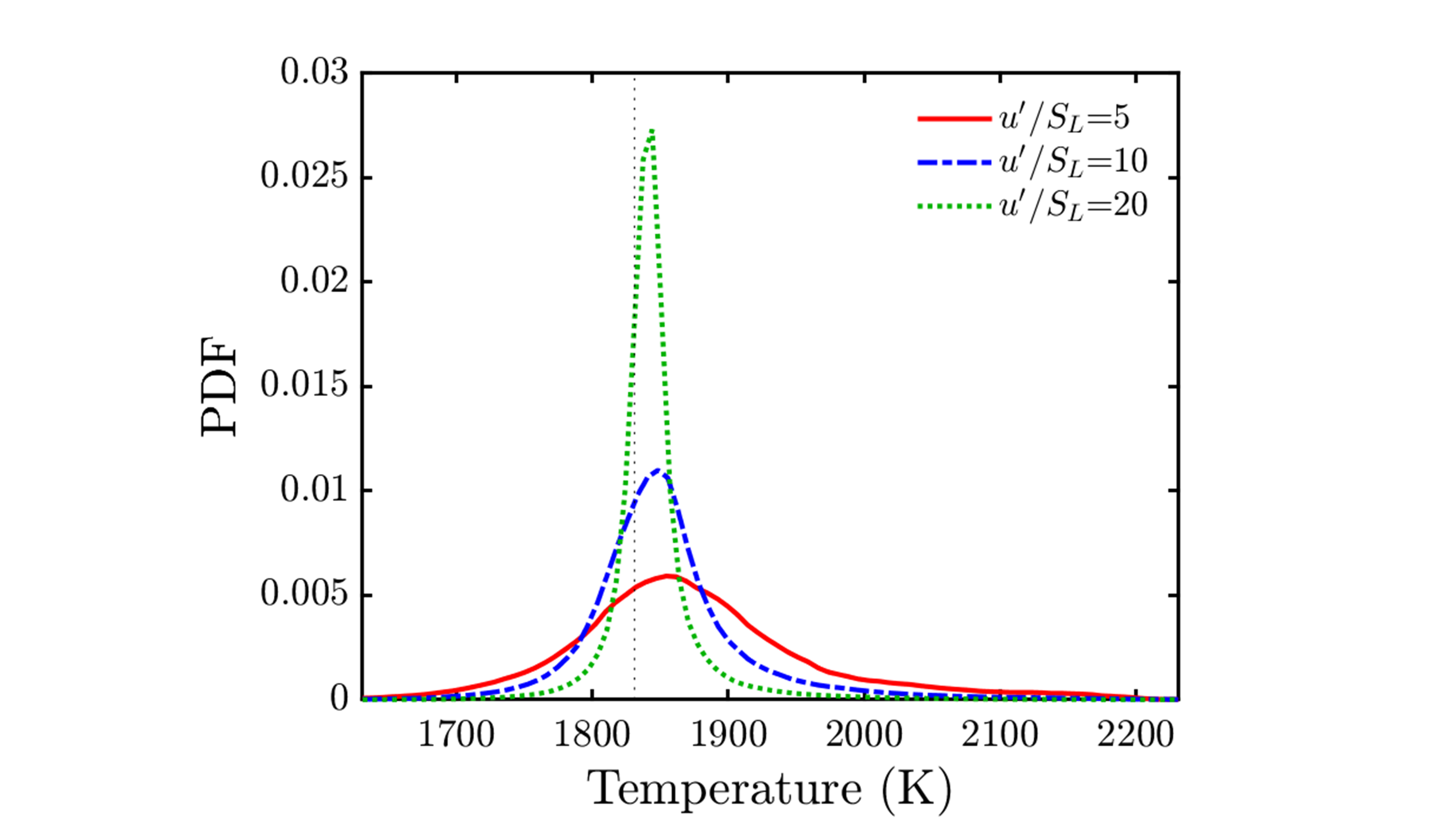}
\caption{\footnotesize Temperature PDF in the post-flame zone at different turbulence intensities. The vertical line represents the adiabatic temperature of 1831~K.}
\vspace{-0.1 in}
\label{fig_tpdf}
\end{figure}

Despite higher local NO production rates within the flame brush at higher $u'$, the global NO production for a given residence time is lower. \highlight{One reason is that} increasing $u'$ raises the turbulent flame speed more rapidly than it increases the flame brush thickness. Indeed, the normalized flame brush thickness $\delta_b/l_f$ is \review{5.13, 5.72, and 7.26} for $u'/S_L = 5$, 10, and 20, respectively, over which the turbulent flame speed increases by approximately a factor of four. \highlight{Another potential reason is the mixing out of high-temperature regions as turbulence intensity increases. Figure~\ref{fig_tpdf} examines the probability density function (PDF) of temperature in a narrow post-flame zone starting at $c > 0.99$ and extending over 0.05~ms of residence time. Clearly, increasing turbulence intensity results in a less stratified post-flame mixture. This indicates that at higher turbulence intensities, the post-flame NO formation rate would actually be reduced due to suppressed super-adiabatic regions.}

Nevertheless, the total NO produced at a given residence time always substantially exceeds either the 1D or 3D laminar value, regardless of the turbulence level. \review{More quantitatively, at $\tau_{res} = 0.1$~ms, the NO mass fractions in the 3D laminar case ($Y_{NO}^{0.1\mathrm{ms}}=3.2 \times 10^{-6}$) and the U05 case ($Y_{NO}^{0.1\mathrm{ms}}=5.2 \times 10^{-6}$) respectively are three and five times the value in the 1D steady flame ($Y_{NO}^{0.1\mathrm{ms}}=1.0 \times 10^{-6}$). The local NO formation rates for $\tau_{res} > 0.1$~ms are nearly constant for all cases, with a value of $\sim 3 \times 10^{-3} s^{-1}$. This would translate to approximately a factor of two difference at a total residence time of 2~ms (corresponding to 20 crank angle degrees at an engine speed of 1500 rpm), a typical timescale in heavy-duty engines during which strong heat release takes place~\cite{Wang2025icef}}. A comparison of Figs.~\ref{fig_YNO_spatialProfile_diffU} and~\ref{fig_YNO_residence_time_Profile_diffU} further emphasizes that meaningful NOx comparisons should be conducted in residence time space rather than in spatial coordinates.

\subsection{NOx reaction pathway analysis} \addvspace{5pt}

\begin{figure*}[ht!]
\centering
\begin{subfigure}[t]{0.9\linewidth}
    \centering
    \includegraphics[trim={1cm 4.5cm 0cm 2cm}, clip=true, width=1\linewidth]{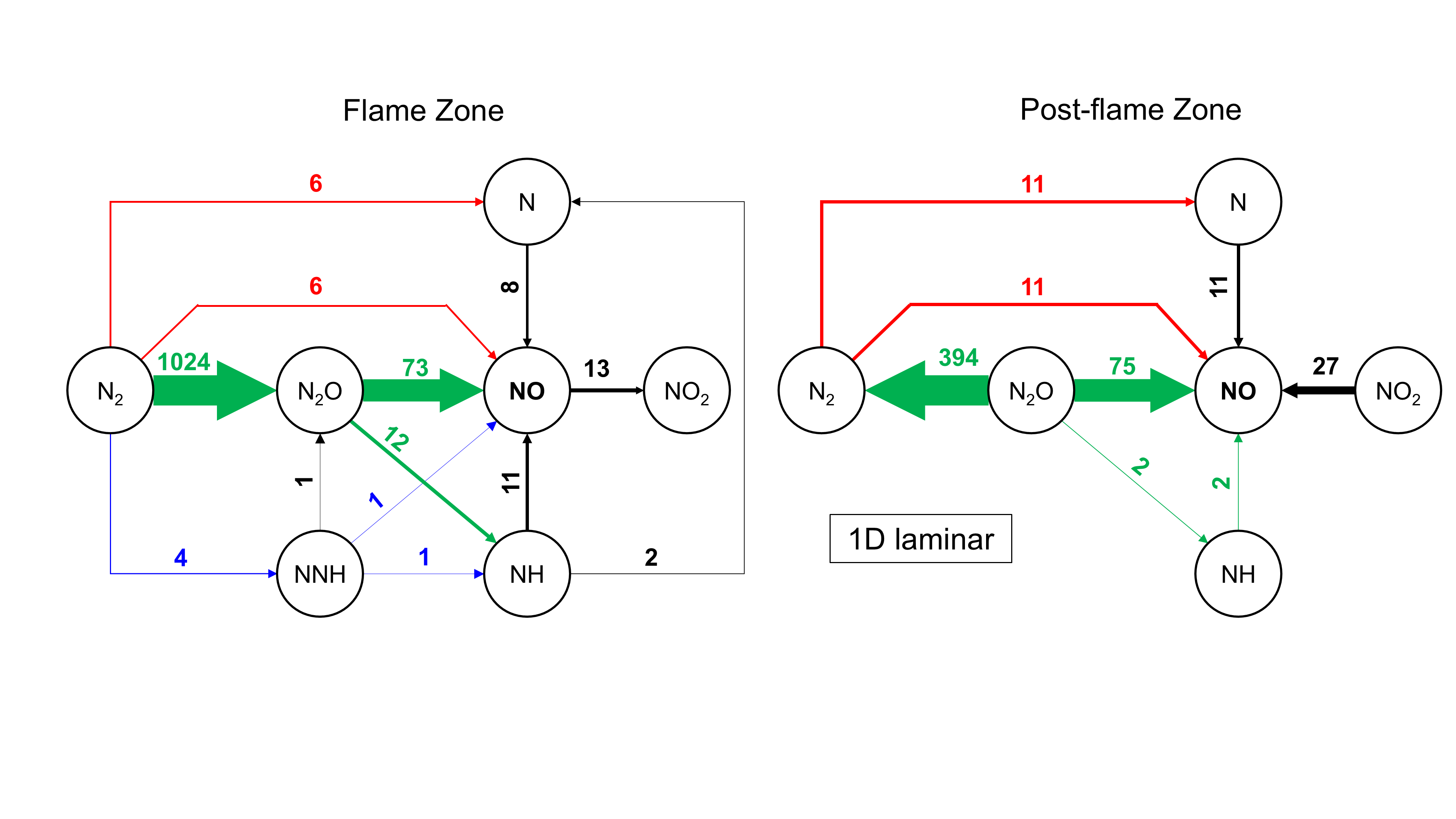}
    \label{fig:pathway_1D}
    \vspace{-0.1 in}
\end{subfigure}
\begin{subfigure}[t]{0.9\linewidth}
    \centering
    \includegraphics[trim={1cm 4.5cm 0cm 3.5cm}, clip=true, width=1\linewidth]{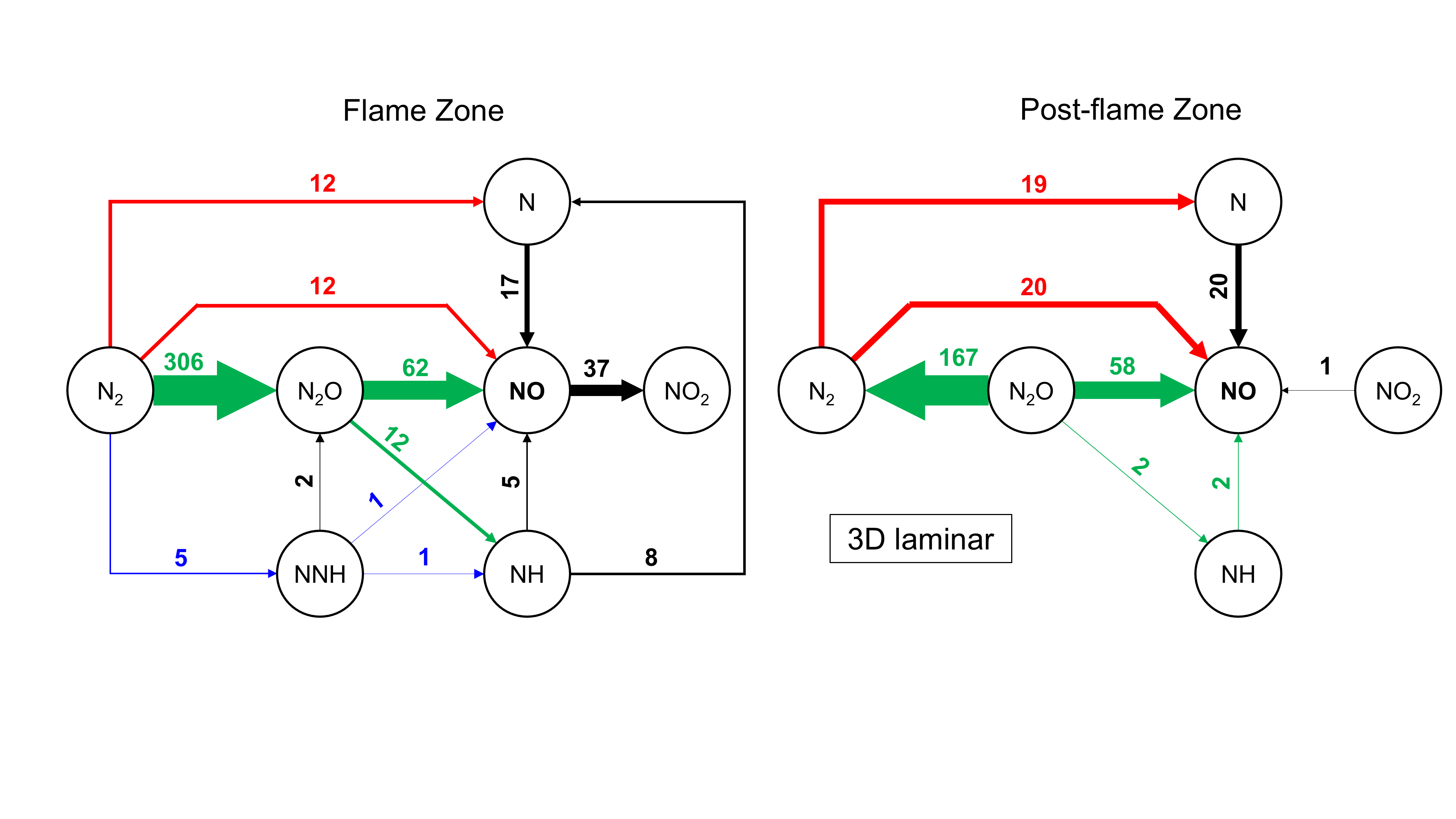}
    \label{fig:pathway_3D}
    \vspace{-0.1 in}
\end{subfigure}
\begin{subfigure}[t]{0.9\linewidth}
    \centering
    \includegraphics[trim={1cm 4.5cm 0cm 3.5cm}, clip=true, width=1\linewidth]{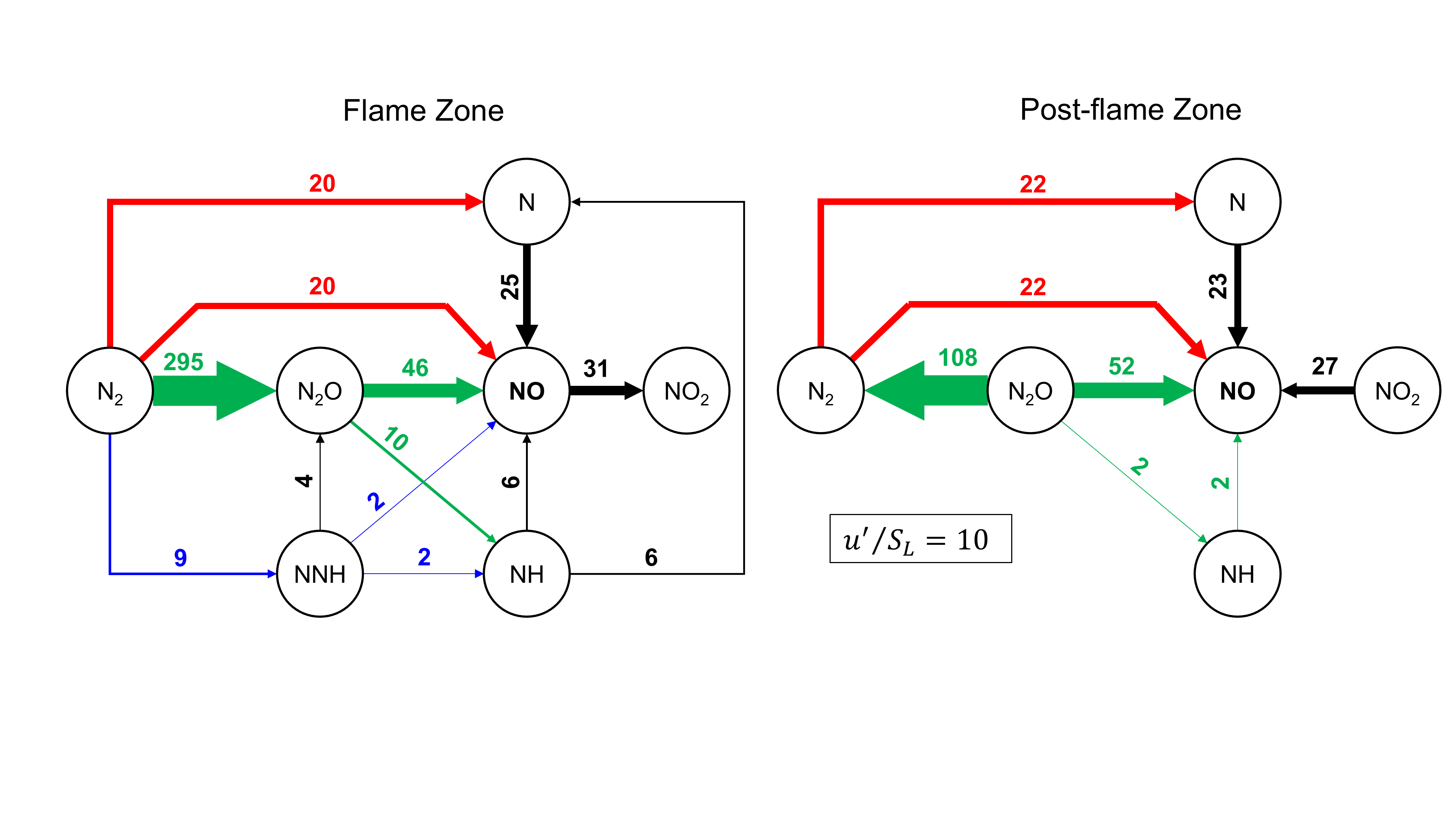}
    \label{fig:pathway_U10}
    \vspace{-0.1 in}
\end{subfigure}
\caption{\footnotesize \review{Reaction pathways of N atom for the flame zone (left) and post-flame zone (right), comparing U10 case with 1D and 3D laminar cases. The arrow width scales with the normalized nitrogen flux but is capped at 100\%. The thermal, \ch{N2O} and NNH pathways are highlighted in red, green, and blue, respectively.} }
\label{fig_pathway}
\vspace{-0.2 in}
\end{figure*}

To further quantify the effect of turbulence on key NO formation mechanisms, a reaction pathway analysis is performed for all the DNS cases, as well as for the 1D and 3D laminar flames, by tracking N atom fluxes between \ch{N2} and NO, for both the flame zone and post-flame zone. The fractional contributions from the three primary NOx pathways (thermal, \ch{N2O}, and NNH) to NO production are quantified and normalized by the total transfer rate of N atoms into NO excluding the contribution from \ch{NO2}. This normalization differs from that used in previous studies~\cite{Day2011,Trisjono2017,Wen2024b}, where individual N atom fluxes were normalized by the total N atom flux leaving \ch{N2}. The present approach enables a more direct quantification of NO formation pathways, rather than \ch{N2} consumption pathways, since not all N atoms leaving \ch{N2} are ultimately converted to NO.

Figure~\ref{fig_pathway} shows the NO reaction pathways for the U10 case in comparison with the 1D and 3D laminar cases. Note that only the pathways with fractional contributions larger than 1\% are retained. Results for the U05 and U20 cases can be found in the supplementary material. As seen, in the 1D steady flame, contributions to flame-zone NO from the three pathways (thermal~:~\ch{N2O}~:~NNH) are 12\%:85\%:2\%, while contributions to post-flame NO are 22\%:77\%:0\%. \review{These ratios change to 24\%:74\%:2\% for the flame zone and 39\%:60\%:0\% for the post-flame zone in the 3D laminar case.} In the turbulent cases, the flame-zone contributions are 36\%:60\%:4\% for U05, 40\%:56\%:4\% for U10, and 44\%:50\%:6\% for U20. It can therefore be concluded that for the lean \ch{H2}/air flames studied here: (1) the \ch{N2O} intermediate is the dominant NO formation mechanism in the 1D steady flame; and (2) as turbulence intensity increases, the thermal and NNH pathways become progressively more important within the flame brush, with the thermal pathway eventually becoming comparable to the \ch{N2O} pathway. This demonstrates that even within the flame zone under turbulent conditions, the thermal pathway can become significant, a finding that has not been widely discussed in the literature.

In the post-flame zone, the fractional contributions are 50\%:49\%:0\% for U05, 44\%:54\%:0\% for U10, and 38\%:61\%:0\% for U20. Clearly, the contribution from the thermal pathway decreases with increasing turbulence intensity, approaching the 3D laminar values. \highlight{This can be explained by Fig.~\ref{fig_tpdf}, which shows that a higher turbulence intensity leads to a more homogeneous} post-flame mixture with a mean temperature approaching the adiabatic flame temperature. This suggests that moderate turbulence promotes post-flame NO formation by sustaining super-adiabatic temperatures that accelerate the thermal pathway, whereas high turbulence suppresses this effect by homogenizing the post-flame mixture.

\subsection{Effects of preferential diffusion} \addvspace{5pt}

\begin{figure*}[h!]
\centering
\includegraphics[trim={2cm 0 2cm 0.1cm}, clip=true, width=0.8\linewidth]{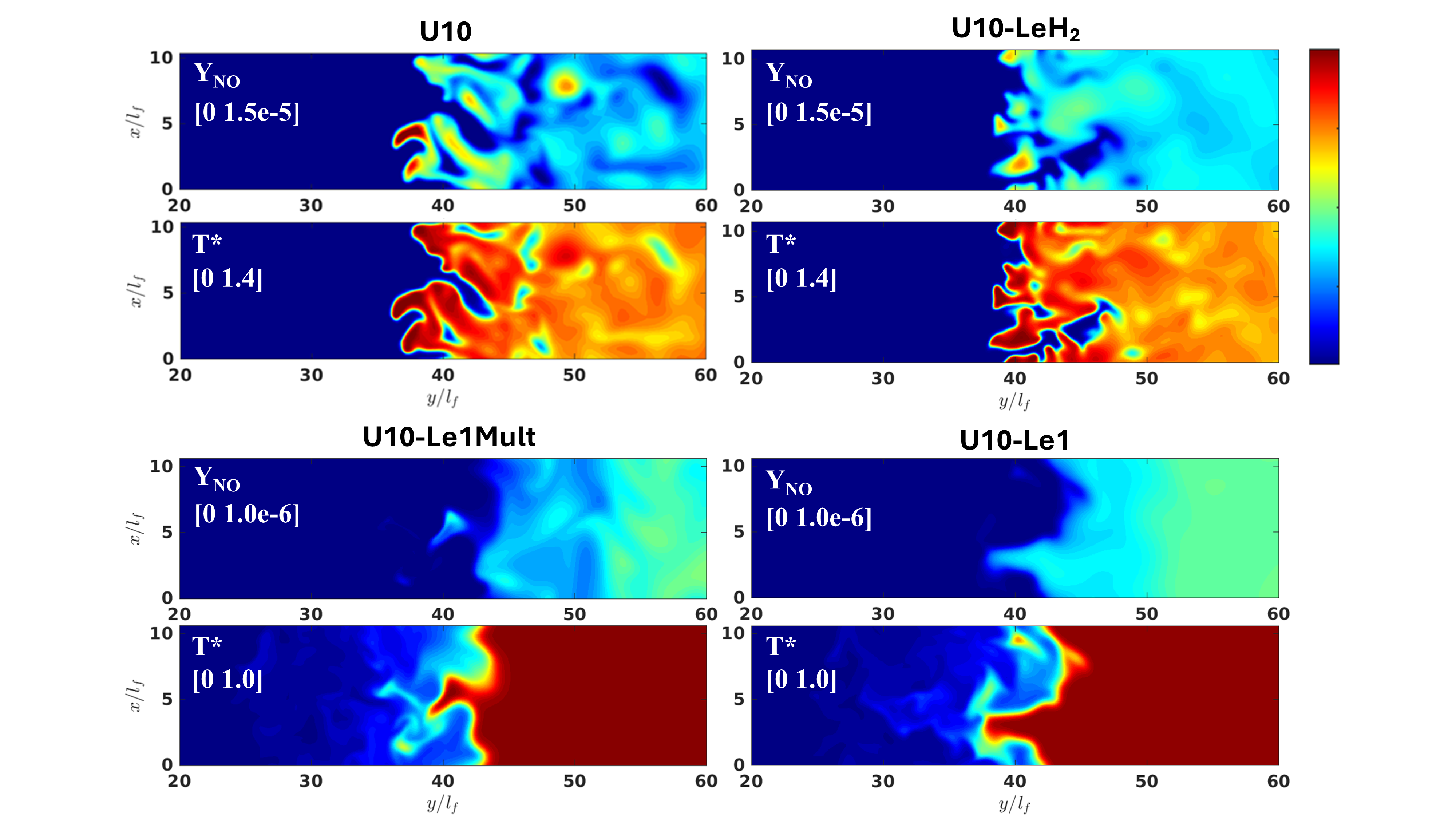}
\caption{\footnotesize Contour plots of $Y_{NO}$ and normalized temperature $T^*$ from DNS cases with different molecular transport models.}
\vspace{-0.2 in}
\label{fig_YNO_2Dcountours_major_diffModel}
\end{figure*}

To better understand the individual roles of the Lewis number and preferential diffusion effects, Fig.~\ref{fig_YNO_2Dcountours_major_diffModel} shows contour plots of NO mass fraction and normalized temperature for different molecular transport models. Eliminating preferential diffusion only (``U10-LeH$_2$'') maintains a flame structure similar to the mixture-averaged case (``U10''), resulting in comparable local NO enhancement. However, in the cases where only Lewis number effects are removed (``U10-Le1Mult'') or both Lewis number and preferential diffusion effects are removed (``U10-Le1''), the flame structure changes drastically: super-adiabaticity is effectively eliminated, and significantly lower NO levels are observed. This suggests that Lewis number effects are the dominant mechanism promoting local NO formation within the flame brush, while preferential diffusion plays a secondary role. In other words, the enhanced NO formation in local hot flame pockets is driven primarily by the sub-unity Lewis number rather than by preferential diffusion, despite the latter also inducing substantial local variations in equivalence ratio.

\begin{figure}[h!]
\centering
\includegraphics[trim={6.2cm 0 6.2cm 0.8cm}, clip=true, width=0.9\linewidth]{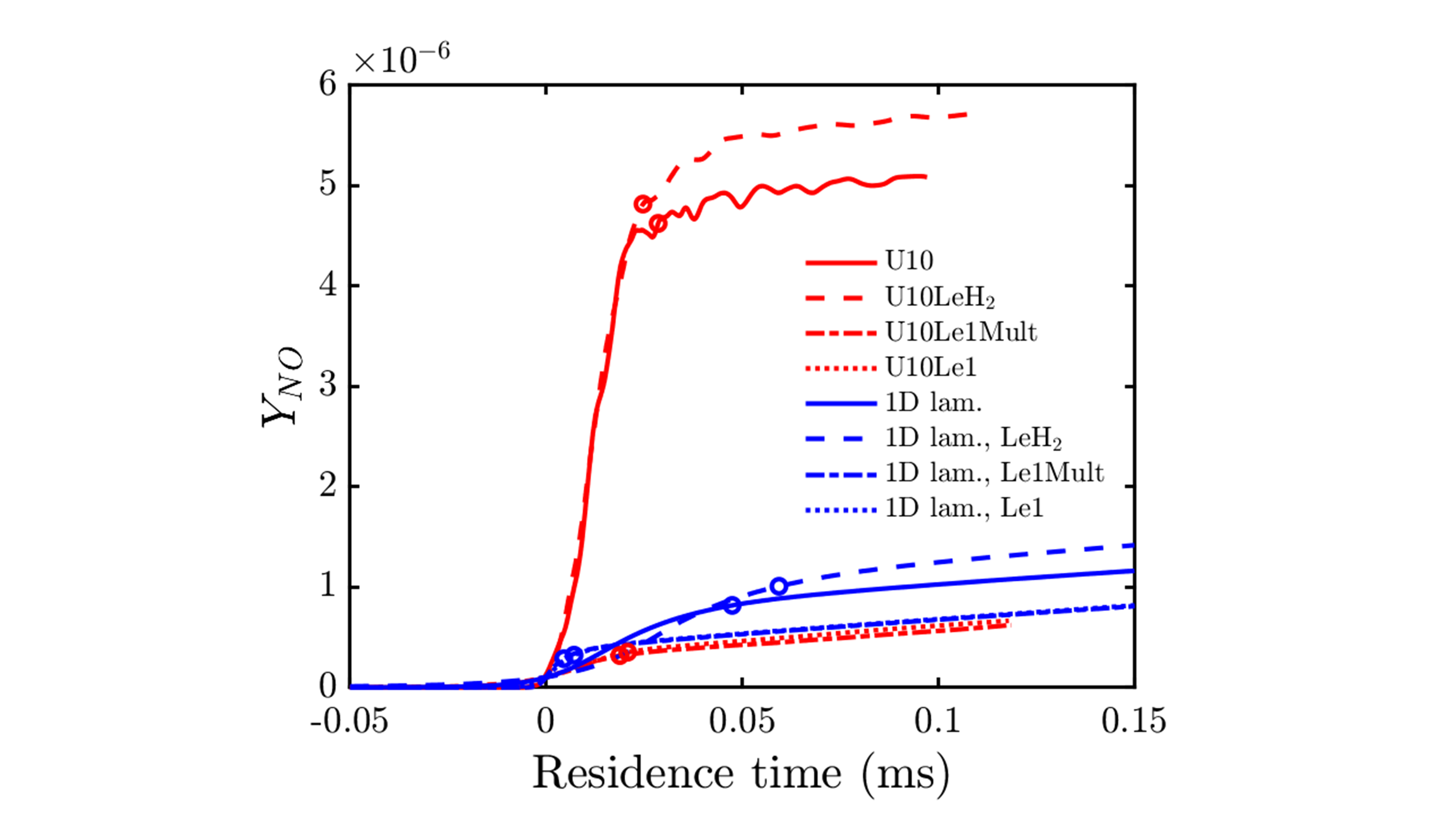}
\caption{\footnotesize Planar- and time-averaged $Y_{NO}$ profiles in the residence time space, with different molecular transport models at $u'/S_L=10$. Time zero and open circles are defined similarly as in Fig.~\ref{fig_YNO_residence_time_Profile_diffU}.}
\vspace{-0.1 in}
\label{fig_YNO_residence_time_Profile_dd_effect}
\end{figure}

Figure~\ref{fig_YNO_residence_time_Profile_dd_effect} shows the NO mass fraction as a function of residence time for $u'/S_L = 10$ with different transport model treatments, along with the corresponding 1D steady flame solutions for comparison. When the fuel Lewis number approaches unity (``U10-Le1Mult'' and ``U10-Le1''), the NO trajectory closely follows the 1D solution, with global NO formation in the turbulent cases being nearly identical to that of the 1D steady flames. In contrast, both non-unity Lewis number turbulent cases (``U10'' and ``U10-LeH$_2$'') exhibit significant NO enhancement across the flame brush. These two cases show an almost identical NO reaction rate in the flame zone while the ``U10-LeH$_2$'' case exhibits a higher post-flame NO mass fraction than ``U10''. This can be attributed to the differences in post-flame NO mass fraction from the corresponding 1D solutions rather than to differential diffusion effects. This further suggests that for unity Lewis number flames, NOx can be modeled without explicitly accounting for turbulence-chemistry interactions. However, for realistic lean \ch{H2} flames where Lewis numbers are significantly below unity, interactions between turbulence and local NOx chemistry must be properly modeled for accurate NOx prediction in practical combustion systems. 

\subsection{Implications for NOx modeling} \addvspace{5pt}

From the analyses above (cf.\ Fig.~\ref{fig_YNO_residence_time_Profile_diffU}), it is clear that the overall NO formation process comprises two distinct stages: a fast NO formation stage within the flame zone, and a slow NO formation stage in the post-flame zone. In the flame-zone stage, all three NO formation pathways are important, and the local NO formation rate is strongly accelerated by turbulence--instability interactions. In the post-flame stage, the local NO formation rate is comparable to the 1D steady flame. These observations have two critical implications for NOx modeling in practical engine simulations. First, all primary NOx formation pathways must be accounted for. Second, the impact of turbulence--chemistry interaction must be properly represented, particularly for the flame-zone stage. Based on this understanding, the global NO formation $Y_{NO}^{tot}$ over a given residence time $\tau_{res}$ can be expressed as
\vspace{-0.05 in}
\begin{equation}
Y_{NO}^{tot} = \overline{\dot{\omega}}_{NO}^{flame}\,\tau_{res}^{flame} + \overline{\dot{\omega}}_{NO}^{post}\,(\tau_{res} - \tau_{res}^{flame})\,,
\vspace{-0.05 in}
\end{equation}
where $\tau_{res}^{flame} \approx \delta_b/S_T$ (see supplementary material) is the residence time within the flame zone, and $\overline{\dot{\omega}}_{NO}^{flame}$ and $\overline{\dot{\omega}}_{NO}^{post}$ are the mean NO reaction rates in the flame zone and the post-flame region, respectively. Based on the observations from Sec.~\ref{sec:turbulenceEffect}, $\overline{\dot{\omega}}_{NO}^{post}$ can be approximated by the corresponding values in 1D steady flames ($\overline{\dot{\omega}}_{NO}^{post,lam}$) as
\vspace{-0.05 in}
\begin{equation}
\overline{\dot{\omega}}_{NO}^{post} \approx \overline{\dot{\omega}}_{NO}^{post,lam}.
\label{eq:flame_rate}
\vspace{-0.05 in}
\end{equation}

\begin{figure}[h!]
\centering
\vspace{-0.05 in}
\includegraphics[trim={5cm -0.2cm 6cm 0.8cm}, clip=true, width=0.9\linewidth]{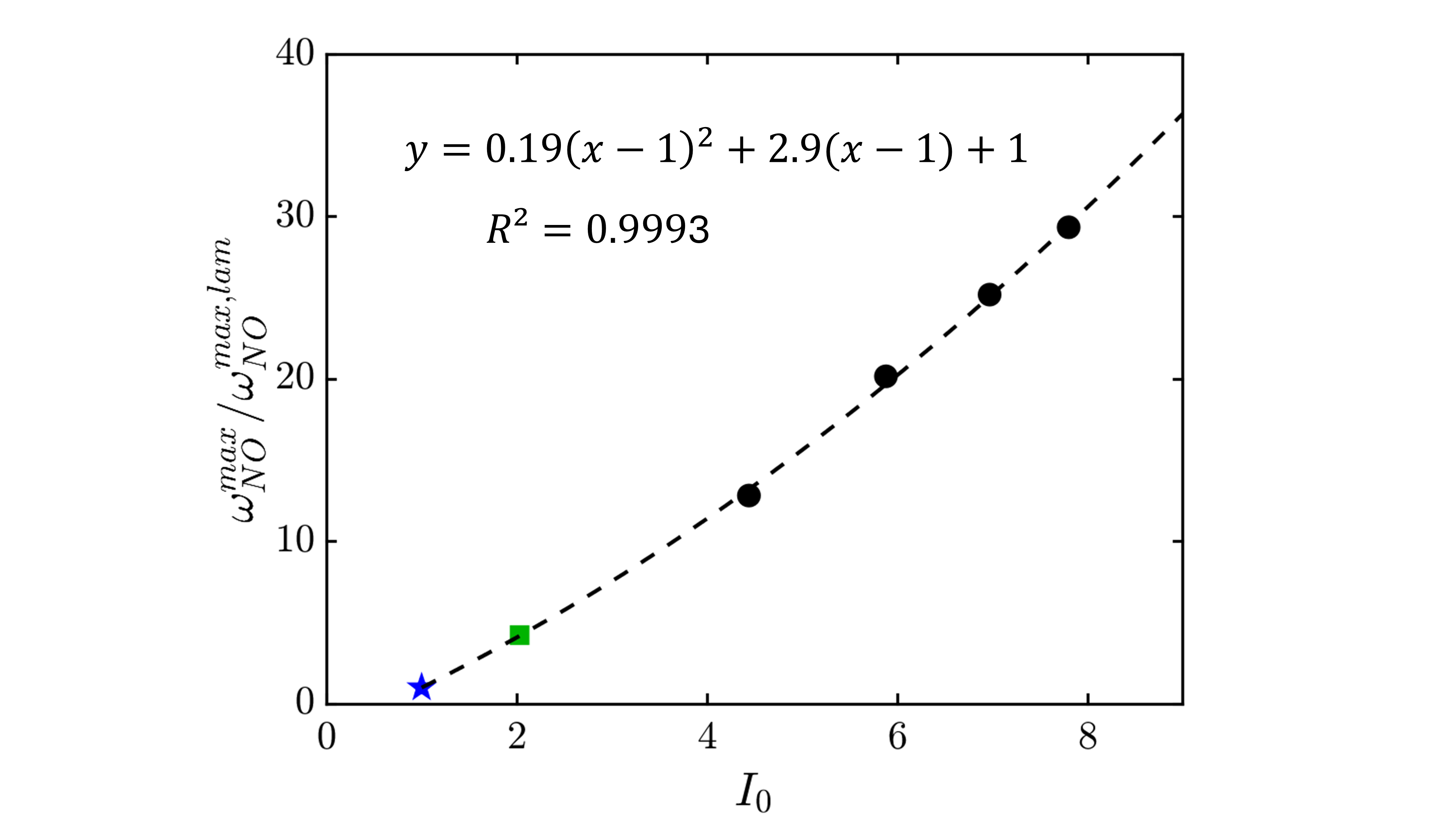}
\caption{\footnotesize Normalized peak conditional mean NO reaction rate as function of $I_0$. Circles indicate turbulent cases, the star indicates the 1D steady case, \review{and the square indicates the 3D laminar case}. Dashed line represents the polynomial fit.}
\vspace{-0.2 in}
\label{fig_wno_correlation}
\end{figure}

Given that the NO reaction rate is positively correlated with both turbulence intensity and radical levels (cf.\ Fig.~\ref{fig_scatter_plot}), which in turn govern the stretch factor $I_0$, the potential correlation between flame-zone NO enhancement and $I_0$ is explored. Figure~\ref{fig_wno_correlation} plots the peak conditional mean NO reaction rate $\omega_{NO}^{max}$ (in progress variable space) normalized by the peak value from the 1D steady flame ($\dot{\omega}_{NO}^{max,lam}$) as a function of $I_0$. An excellent correlation is obtained with an $R^2$ value greater than 0.999 when the data are fitted to a second-order polynomial. Specifically,
\vspace{-0.05 in}
\begin{equation}
f = 0.19\,(I_0 - 1)^2 + 2.9\,(I_0 - 1) + 1\,,
\vspace{-0.05 in}
\end{equation}
where $f = \dot{\omega}_{NO}^{max}/\dot{\omega}_{NO}^{max,lam}$. While further data are needed to confirm the applicability of this correlation to a wider range of conditions, this result is encouraging and provides the basis for constructing improved NOx models for practical engine simulations. It is now possible to model $\overline{\dot{\omega}}_{NO}^{flame}$ as
\vspace{-0.05 in}
\begin{equation}
\overline{\dot{\omega}}_{NO}^{flame} = f \cdot \overline{\dot{\omega}}_{NO}^{flame,lam}\,.
\label{eq:post_rate}
\vspace{-0.05 in}
\end{equation}
\review{where $\overline{\dot{\omega}}_{NO}^{flame,lam}$ is the flame-zone NO reaction rate in 1D steady flames.}

Equations~\ref{eq:flame_rate} and \ref{eq:post_rate} can be used in practical \review{RANS-based} engine simulations to compute the local NO formation rates in the different flame zones, using turbulent flame quantities already available from the combustion model (e.g., $I_0$ and $\delta_b/S_T$) and laminar quantities that can be conveniently tabulated using 1D unstretched steady flames.

\section{Conclusions\label{sec:unnum}} \addvspace{5pt}

This work presented DNS results for lean premixed hydrogen--air flames under engine-relevant conditions to investigate NOx formation across varying turbulence intensities and molecular transport models. It was shown that the global NO production was significantly enhanced in all of the turbulent cases relative to the 1D steady flame, reaching five times the 1D value at a residence time of 0.1~ms. Increasing turbulence intensity was found to promote the NO formation rate within the flame brush by inducing super-adiabatic hot spots and elevating key flame radical concentrations, while simultaneously reducing the flame-brush residence time and suppressing \highlight{super-adiabatic regions} in the post-flame zone, resulting in lower overall NO levels at higher turbulence intensities. By systematically varying the molecular transport model, Lewis number effects were identified as the primary driver of NO enhancement, with preferential diffusion playing only a secondary role. Finally, a novel correlation between the flame-zone NO production rate and the stretch factor was established, based on which a conceptual model for predicting the NO enhancement in RANS simulations was proposed. These findings provide important physical insights into turbulence--NOx chemistry interactions and lay the groundwork for improved NOx modeling under ultra-lean, engine-relevant conditions.

\acknowledgement{CRediT authorship contribution statement} \addvspace{5pt}

{\bf Chao Xu}: Conceptualization, Methodology, Data curation, Formal analysis, Writing – original draft, Funding acquisition. {\bf Yiqing Wang}: Methodology, Data curation, Formal analysis, Writing – review and editing. {\bf Riccardo Scarcelli}: Writing – review and editing, Funding acquisition.

\acknowledgement{Declaration of competing interest} \addvspace{5pt}

The authors declare that they have no known competing financial interests or personal relationships that could have appeared to influence this work.

\acknowledgement{Acknowledgments} \addvspace{5pt}

The submitted manuscript has been created by UChicago Argonne, LLC, Operator of Argonne National Laboratory (“Argonne”). Argonne, a U.S. Department of Energy Office of Science laboratory, is operated under Contract No. DE-AC02-06CH11357. The U.S. Government retains for itself, and others acting on its behalf, a paid-up nonexclusive, irrevocable worldwide license in said article to reproduce, prepare derivative works, distribute copies to the public, and perform publicly and display publicly, by or on behalf of the Government. The Department of Energy will provide public access to these results of federally sponsored research in accordance with the DOE Public Access Plan (\url{http://energy.gov/downloads/doe-public-access-plan}).

The authors gratefully acknowledge Gurpreet Singh, Program Manager at the Transportation Technologies Office, U.S. Department of Energy's Office of Critical Minerals and Energy Innovation, for supporting this work. The authors also acknowledge the Laboratory Computing Resource Center (LCRC) at Argonne National Laboratory for providing the computing resources.

\footnotesize
\baselineskip 9pt

\clearpage
\thispagestyle{empty}
\bibliographystyle{proci}
\bibliography{PROCI_LaTeX}


\newpage

\small
\baselineskip 10pt


\end{document}